%% file: main.tex
\documentclass[12pt]{article}
\usepackage{graphicx}
\usepackage{mathptmx}
\usepackage[intlimits,centertags]{amsmath}
\usepackage{cite}
\usepackage{amssymb,amsfonts}
\usepackage[pdftex]{hyperref}
\usepackage{aas_macros}
\usepackage{enumerate}
\usepackage{xfrac}
\usepackage[table,xcdraw]{xcolor}
\usepackage{sectsty}
\usepackage[normalem]{ulem}
\usepackage{setspace}
\usepackage[verbose]{wrapfig}
\usepackage{ragged2e}
\usepackage{etoolbox} 
\usepackage{combelow}
\usepackage{lineno}
\usepackage[symbol]{footmisc}
\usepackage[T1]{fontenc}
\usepackage{pdfpages}
\usepackage{times}
\usepackage[compact]{titlesec}
\usepackage{geometry}
\usepackage[utf8]{inputenc}
\usepackage{enumitem,amssymb}
\usepackage{ragged2e}
\newlist{thematic}{itemize}{8}
\setlist[thematic]{label=$\square$}
\usepackage{pifont}

\newif\ifastrophysical
\astrophysicalfalse
\newcommand{\onlyastrophysical}[1] 
{
  \ifastrophysical
  #1 
  \fi
}

\newif\iffundamental
\fundamentaltrue
\newcommand{\onlyfundamental}[1] 
{
  \iffundamental
  #1 
  \fi
}

\definecolor{header_color}{HTML}{15588c}
\subsectionfont{\color{header_color}}

\begin{document}
\begin{titlepage}
\includepdf{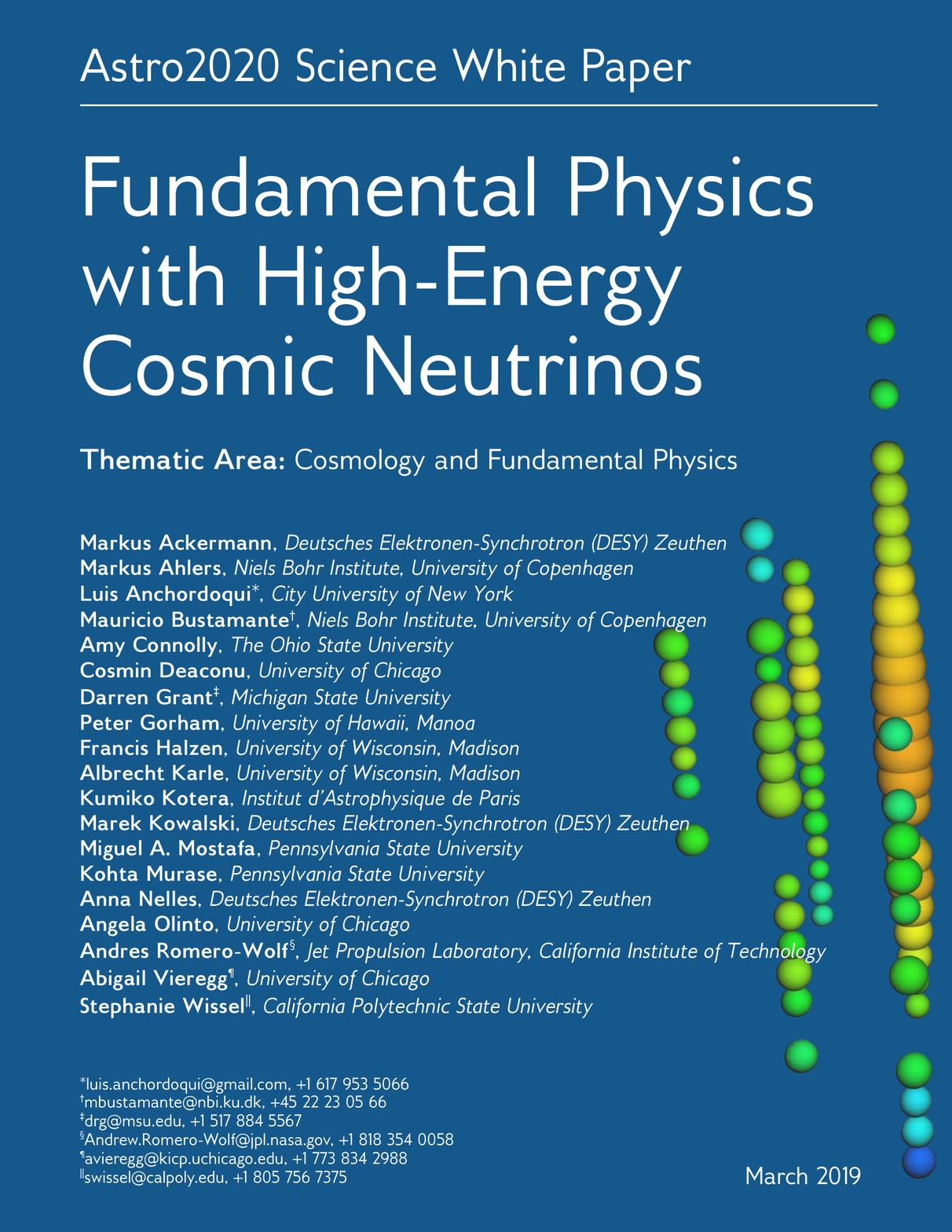}
\end{titlepage}

\pagenumbering{roman}

\begin{center}
\textbf{Abstract}
\end{center}
\justify
High-energy cosmic neutrinos can reveal new fundamental particles and interactions, probing energy and distance scales far exceeding those accessible in the laboratory.  This white paper describes the outstanding particle physics questions that high-energy cosmic neutrinos can address in the coming decade.  A companion white paper discusses how the observation of cosmic neutrinos can address open questions in astrophysics.  Tests of fundamental physics using high-energy cosmic neutrinos will be enabled by detailed measurements of their energy spectrum, arrival directions, flavor composition, and timing.\\

\begin{center}
\textbf{Endorsers}
\linebreak
\input{endorsers} 
\end{center}

\normalsize
\pagebreak

\subsection*{High-Energy Cosmic Neutrinos}

\pagenumbering{arabic}

What are the fundamental particles and interactions of Nature?  High-energy cosmic neutrinos are uniquely poised to explore them in an uncharted and otherwise unreachable energy and distance regime.  They allow us to explore the cosmic and energy frontiers of particle physics, complementing current and future colliders that will explore the energy and intensity frontiers. 

Despite the spectacular success of the Standard Model (SM) of particle physics, we know that it must be extended to account for at least the existence of neutrino mass, dark matter, and dark energy.
A common feature of many theories beyond the Standard Model (BSM) is that their effects are more clearly apparent the higher the energy of the process, where new particles, interactions, and symmetries, undetectable at lower energies, could make themselves evident.  Yet, particle colliders have failed to find clear evidence of BSM physics up to TeV energies, the highest reachable in the lab. To access particle interactions beyond the TeV scale, we must use particle beams made by natural cosmic accelerators.  They produce the most energetic neutrinos, photons, and charged particles known, with energies orders of magnitude higher than in man-made colliders.   

\begin{figure}[t!]
  \vspace*{-0.5cm}
  \centering
  \includegraphics[width=0.96\textwidth]{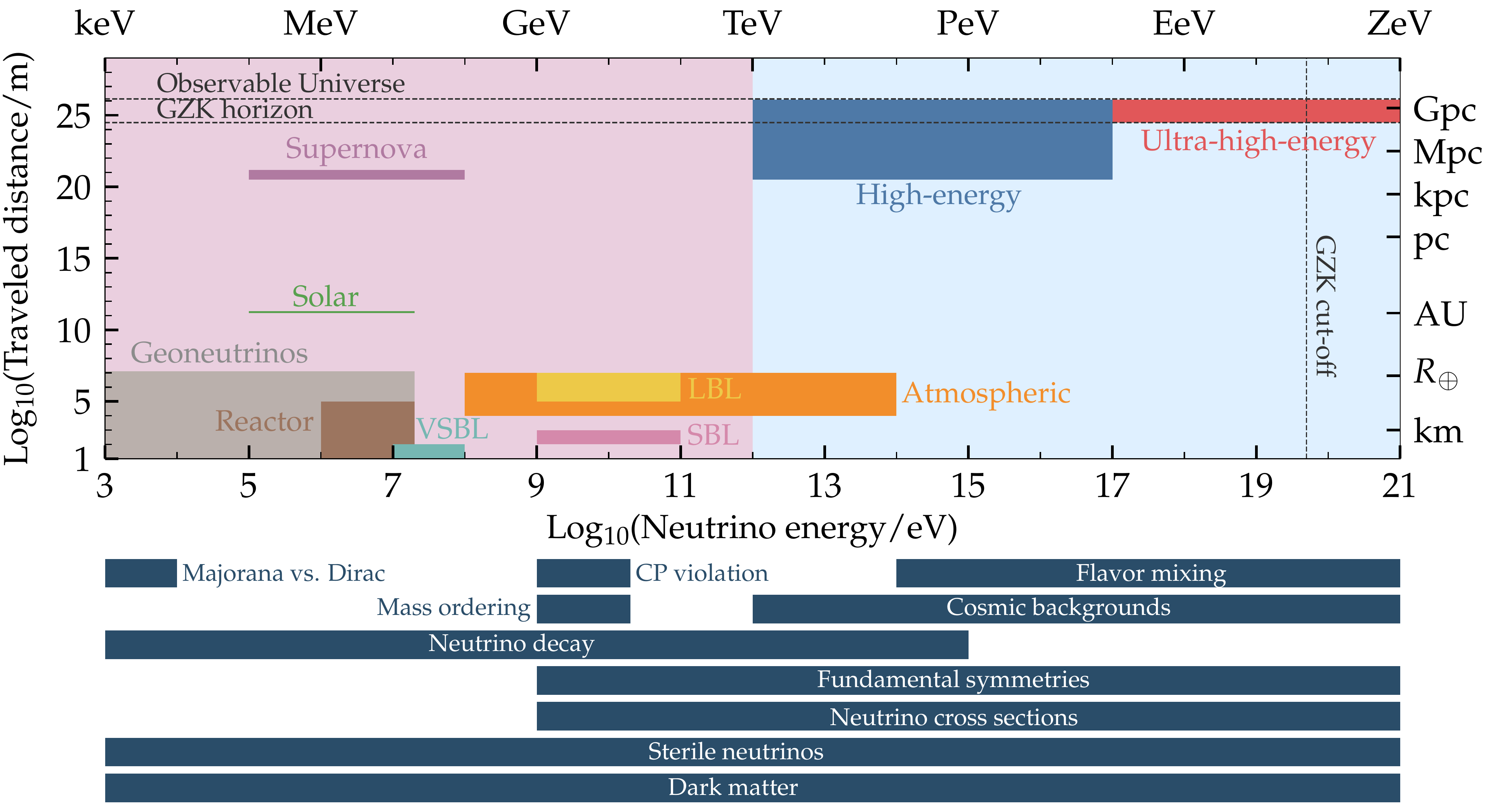}
  \caption{\label{fig:scales}Tests of fundamental physics accessible with neutrinos of different energies.}
\end{figure}

Cosmic neutrinos are especially fitting probes of fundamental physics beyond the TeV scale, as shown in Fig.\ \ref{fig:scales}.  First, cosmic neutrinos reach higher energies than neutrinos made in the Sun, supernovae, the atmosphere of Earth, particle accelerators, and nuclear reactors. Further, they reach Earth with energies higher than that of gamma rays and likely as high as ultra-high-energy (UHE) cosmic rays.  Second, because most cosmic neutrinos come from extragalactic sources located at cosmological distances, even tiny BSM effects could accumulate up to observable levels as neutrinos travel to Earth, having crossed essentially the observable Universe.  And, third, because the propagation of neutrinos from the sources to 
the detectors is well understood and predicted by the SM, BSM effects could be more easily spotted than in charged particles.  

Tests of fundamental physics using cosmic neutrinos are possible in spite of  astrophysical and cosmological uncertainties.  Yet this endeavor is not without challenges: the neutrino detection cross section is tiny and cosmic neutrino fluxes are expected to fall rapidly with neutrino energy.  Nevertheless, we show below that these obstacles are either surmountable or can be planned for.  

\subsection*{Open Questions: What Can High-Energy Cosmic Neutrinos Test?}

Figure \ref{fig:scales} shows the wide breadth of important open questions in fundamental physics that cosmic neutrinos can address\ \cite{Ahlers:2018mkf,Anchordoqui:2005is,Marfatia:2015hva}.  They complement questions tackled by neutrinos of lower energies.  

Cosmic neutrinos span a wide range in energy.  In the TeV--PeV range, astrophysical neutrinos are regularly detected by IceCube\ \cite{Aartsen:2013bka,Razzaque:2013uoa,Aartsen:2013jdh,Aartsen:2014gkd,Aartsen:2015rwa,Aartsen:2016xlq} from what are likely mainly extragalactic sources\ \cite{Ahlers:2013xia,Anchordoqui:2013dnh,Ahlers:2015moa,Denton:2017csz,Aartsen:2017ujz,IceCube:2018dnn,IceCube:2018cha}.  At the EeV scale, cosmogenic neutrinos, produced by UHE cosmic rays interacting with photon backgrounds through the GZK effect\ \cite{Greisen:1966jv, Zatsepin:1966jv}, are predicted but have not yet been observed\ \cite{Zas:2017xdj, Aartsen:2018vtx, Gorham:2019guw}.  See Ref.\ \cite{Ackermann:2019ows} for a discussion of astrophysics enabled by observations of cosmic neutrinos.

\smallskip

\begin{figure}[t!]
 %\vspace*{-0.2cm}
 \begin{minipage}[t]{0.492\textwidth}  
  \centering
  \begin{minipage}[c][8cm][c]{\linewidth}
  \includegraphics[width=\linewidth]{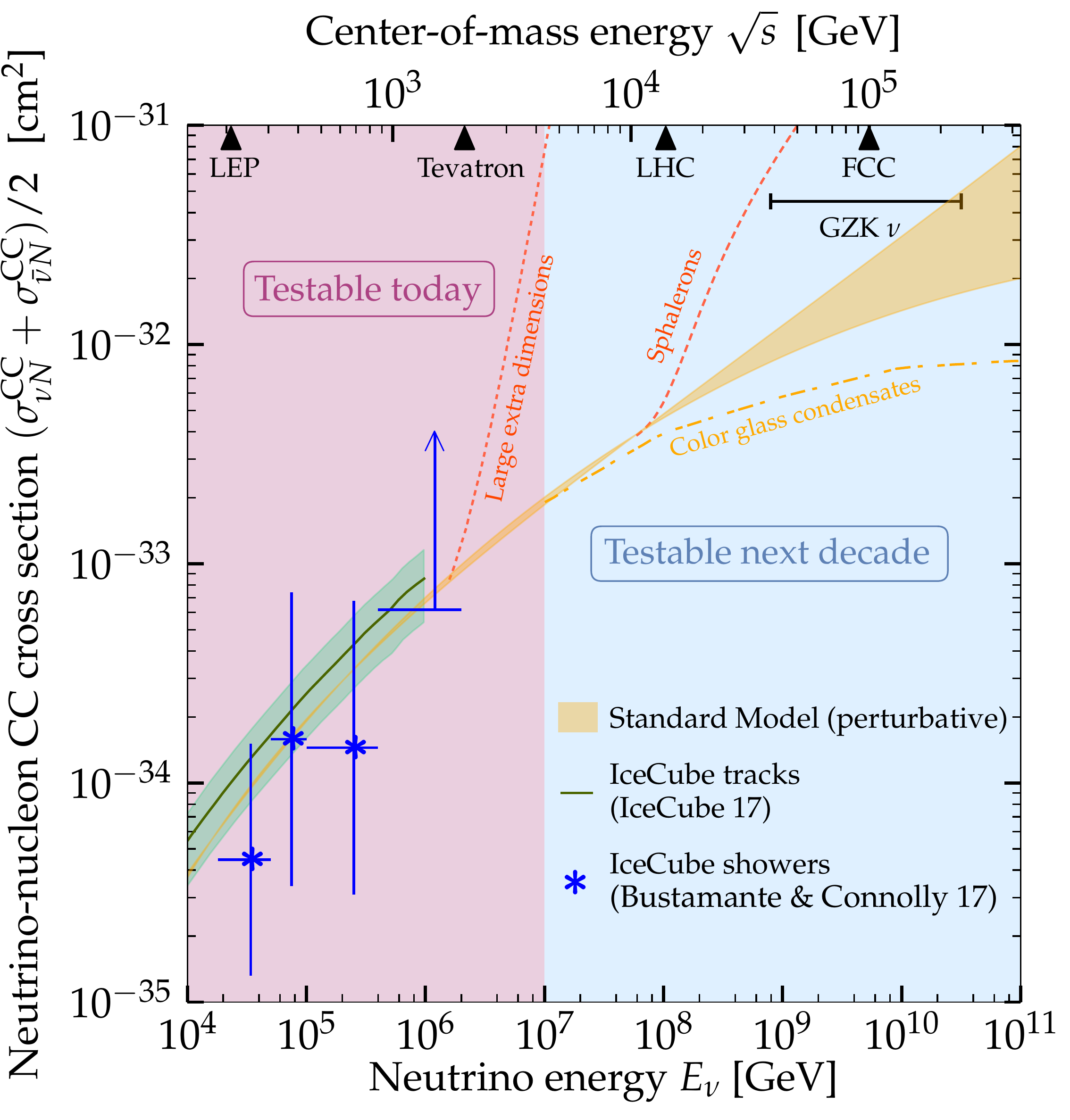}
  \end{minipage}
  \caption{\label{fig:cross_section} Neutrino-nucleon cross section.  Below 1~PeV, measurements\ \cite{Aartsen:2017kpd, Bustamante:2017xuy} are compared to the SM uncertainty band\ \cite{Connolly:2011vc} (see also Ref.\ \cite{CooperSarkar:2011pa}) that encloses predictions\ \cite{Gandhi:1998ri, Connolly:2011vc, CooperSarkar:2011pa, Block:2014kza, Arguelles:2015wba}.  The cross section may change due to new physics --- {\it e.g.}, large extra dimensions\ \cite{AlvarezMuniz:2001mk} (TeV-scale, in tension with LHC results), electroweak sphalerons\ \cite{Ellis:2016dgb} (9-TeV barrier height) --- or non-perturbative effects --- {\it e.g.}, color glass condensate\ \cite{Henley:2005ms} (model BGBK$_{\rm III}$).}
 \end{minipage}
 \hfill
 \begin{minipage}[t]{0.492\textwidth}  
  \centering
  \begin{minipage}[c][8cm][c]{\linewidth}
  \includegraphics[width=\linewidth]{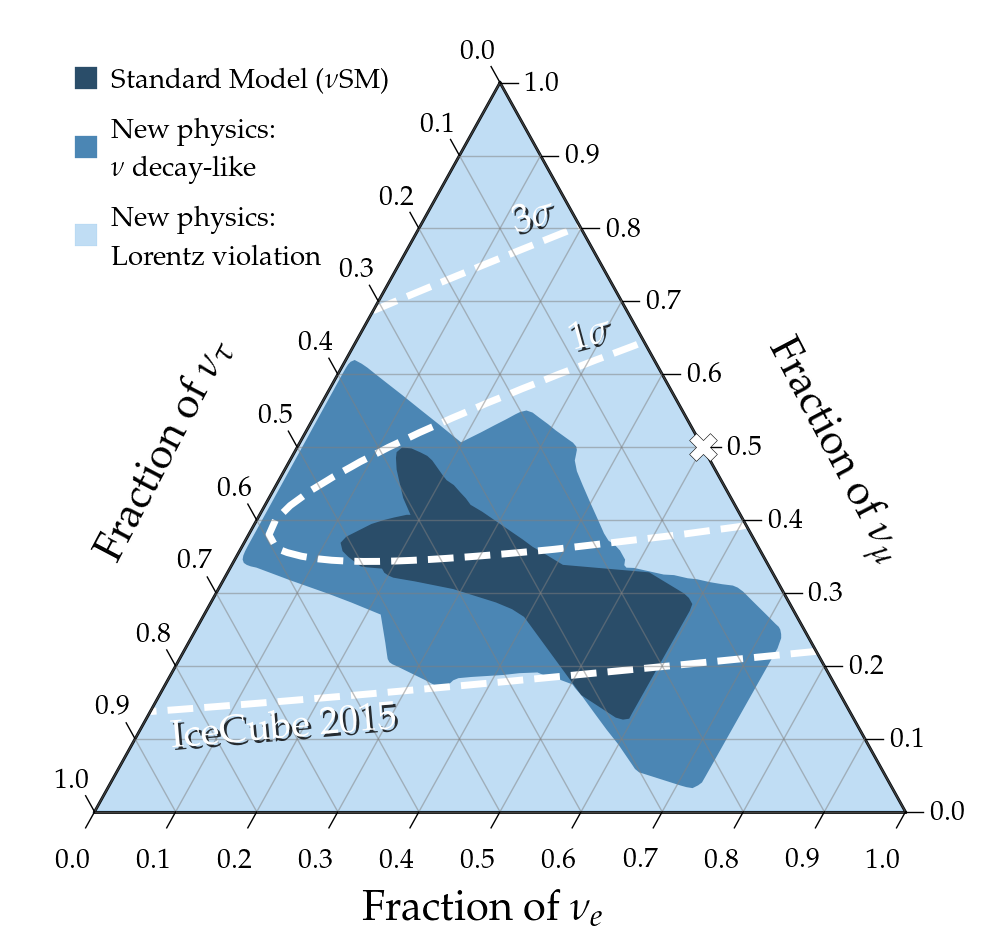}
  \end{minipage}
  \caption{\label{fig:flavor_ratios}Flavor composition at Earth of high-energy cosmic neutrinos, indicating the ``theoretically palatable"\ \cite{Bustamante:2015waa} regions accessible with the Standard Model with massive neutrinos ($\nu$SM), with new physics similar to neutrino decay, and with new physics similar to Lorentz-invariance violation.  The neutrino mixing parameters are generously varied within their uncertainties at $3\sigma$.  The tilt of the tick marks indicates the orientation along which to read the flavor content.}
  \label{fig:parameter_fit}
 \end{minipage}
\end{figure}

\textbf{How do neutrino cross sections behave at high energies?}  The neutrino-nucleon cross section in the TeV--PeV range was measured for the first time using astrophysical and atmospheric neutrinos\ \cite{Aartsen:2017kpd,Bustamante:2017xuy,Aartsen:2018vez}, extending\ \cite{Berezinsky:1974kz, Hooper:2002yq, Hussain:2006wg, Borriello:2007cs, Hussain:2007ba} measurements that used GeV neutrinos from accelerators\ \cite{Conrad:1997ne, Formaggio:2013kya, Tanabashi:2018oca}.  Fig.\ \ref{fig:cross_section} shows that the measurements agree with high-precision SM predictions\ \cite{CooperSarkar:2011pa}. Future measurements in the EeV range would probe BSM modifications of the cross section at center-of-momentum energies of 100~TeV~\cite{Kusenko:2001gj,AlvarezMuniz:2001mk,Anchordoqui:2001cg,Cornet:2001gy,Kowalski:2002gb,AlvarezMuniz:2002ga,Anchordoqui:2005pn,Marfatia:2015hva,Ellis:2016dgb,Anchordoqui:2018qom} and test the structure of nucleons\ \cite{Henley:2005ms, Armesto:2007tg, Enberg:2008te, Illarionov:2011wc, Bhattacharya:2015jpa, Garzelli:2015psa, Halzen:2016pwl, Halzen:2016thi, Bhattacharya:2016jce, Benzke:2017yjn, Giannini:2018utr} more deeply than colliders\ \cite{Anchordoqui:2006ta, Bertone:2018dse}. 

\smallskip\pagebreak

\textbf{How do flavors mix at high energies?}  Experiments with neutrinos of up to TeV energies have confirmed that the different neutrino flavors, $\nu_e$, $\nu_\mu$, and $\nu_\tau$, mix and oscillate into each other as they propagate \cite{Tanabashi:2018oca}.   
Figure\ \ref{fig:flavor_ratios} shows that, if high-energy cosmic neutrinos en route to Earth oscillate as expected, the predicted allowed region of the ratios of each flavor to the total flux is small, even after accounting for uncertainties in the parameters that drive the oscillations and in the neutrino production process\ \cite{Bustamante:2015waa}.  
However, at these energies and over cosmological propagation baselines~\cite{Learned:1994wg}, mixing is untested; BSM effects could affect oscillations, vastly expanding the allowed region of flavor ratios and making them sensitive probes of BSM\ \cite{Beacom:2003nh, Pakvasa:2007dc, Bustamante:2010bf, Bustamante:2010nq, Mehta:2011qb, Bustamante:2015waa, Arguelles:2015dca, Shoemaker:2015qul,Gonzalez-Garcia:2016gpq, Rasmussen:2017ert, Ahlers:2018yom}. 

\smallskip

\textbf{What are the fundamental symmetries of Nature?}  Beyond the TeV scale, the symmetries of the SM may break or new ones may appear.  The effects of breaking lepton-number conservation, or CPT and Lorentz invariance\ \cite{Colladay:1998fq}, cornerstones of the SM, are expected to grow with neutrino energy and affect multiple neutrino observables~\cite{Kostelecky:2003cr, Hooper:2005jp,Kostelecky:2008ts, Kostelecky:2011gq, Gorham:2012qs, Borriello:2013ala, Stecker:2014xja, Anchordoqui:2014hua, Tomar:2015fha, Amelino-Camelia:2015nqa, Liao:2017yuy, Anchordoqui:2006wc}.  Currently, the strongest constraints in neutrinos come from high-energy atmospheric neutrinos\ \cite{Aartsen:2017ibm}; cosmic neutrinos could provide unprecedented sensitivity \cite{AmelinoCamelia:1997gz, Hooper:2005jp, GonzalezGarcia:2005xw, Anchordoqui:2005gj, Bazo:2009en, Bustamante:2010nq, Kostelecky:2011gq, Diaz:2013wia, Stecker:2014oxa, Stecker:2014xja, Tomar:2015fha, Ellis:2018ogq, Laha:2018hsh}.  Further, detection of ZeV neutrinos, well beyond astrophysical expectations, would probe Grand Unified Theories~\cite{Sigl:1996gm, Berezinsky:2009xf, Berezinsky:2011cp, Lunardini:2012ct, Anchordoqui:2018qom}.

\smallskip

\textbf{Are neutrinos stable?}  Neutrinos are essentially stable in the SM\ \cite{Pal:1981rm, Hosotani:1981mq, Nieves:1983fk}, but BSM physics could introduce new channels for the heavier neutrinos to decay into the lighter ones\ \cite{Chikashige:1980qk, Gelmini:1982rr, Tomas:2001dh}, with shorter lifetimes.  During propagation over cosmological baselines, neutrino decay could leave imprints on the energy spectrum and flavor composition\ \cite{Beacom:2002vi, Baerwald:2012kc, Shoemaker:2015qul, Bustamante:2016ciw, Denton:2018aml}.  The associated sensitivity outperforms existing limits obtained using neutrinos with shorter baselines\ \cite{Bustamante:2016ciw}.  Comparable sensitivities are expected for similar BSM models, like pseudo-Dirac neutrinos\ \cite{Beacom:2003eu, Joshipura:2013yba, Shoemaker:2015qul}.

\smallskip

\textbf{What is dark matter?}  Cosmic neutrinos can probe the nature of dark matter.  Dark matter may decay or self-annihilate into neutrinos\ \cite{Feng:2010gw,Beacom:2006tt,Yuksel:2007ac,Murase:2012xs}, leaving imprints on the neutrino energy spectrum, {\it e.g.}, line-like features. 
Searches for these features have yielded strong constraints on dark matter in the Milky Way\ \cite{Adrian-Martinez:2015wey, Aartsen:2016pfc, Aartsen:2017ulx} and nearby galaxies\ \cite{Aartsen:2013dxa}.  
High-energy cosmic neutrinos can probe superheavy dark matter with PeV masses\ \cite{Feldstein:2013kka, Esmaili:2013gha, Higaki:2014dwa, Rott:2014kfa, Dudas:2014bca, Ema:2013nda, Zavala:2014dla, Murase:2015gea, Anchordoqui:2015lqa, Boucenna:2015tra, Dev:2016qbd, Hiroshima:2017hmy, Chianese:2017nwe} and light dark matter\ \cite{Higaki:2014dwa, Fong:2014bsa, Cohen:2016uyg, Hiroshima:2017hmy, Sui:2018bbh}.
Multi-messenger constraints are crucial to assess dark matter explanations of the observed neutrino spectrum\ \cite{Ahlers:2013xia , Bhattacharya:2014vwa, Murase:2015gea, Cohen:2016uyg, Bhattacharya:2017jaw}. 
Further, anisotropies in the neutrino sky towards the Galactic Center can reveal dark matter decaying\ \cite{Bai:2013nga} or interacting with neutrinos\ \cite{Arguelles:2017atb}.

\smallskip

\textbf{Are there hidden interactions with cosmic backgrounds?}  High-energy cosmic neutrinos may interact with low-energy relic neutrino backgrounds via new interactions\ \cite{Lykken:2007kp, Ioka:2014kca, Ng:2014pca, Blum:2014ewa, Shoemaker:2015qul, Altmannshofer:2016brv, Barenboim:2019tux}, with large-scale distributions of matter via new forces\ \cite{Bustamante:2018mzu}, or with dark backgrounds\ \cite{Capozzi:2018bps}, including dark energy\ \cite{Anchordoqui:2007iw,Klop:2017dim}.  These interactions may mimic the existence of neutrino mass, affect the neutrino flavor composition, and induce anisotropies in the high-energy neutrino sky.

%%%%%%%%%%%%%%%%%%%%%%%%%%%%
%%%%%%%%%%%%%%%%%%%%%%%%%%%%

\subsection*{Neutrino Observables: What Do We Use to Probe Fundamental Physics?}

To probe fundamental physics, we look at four neutrino observables, individually or together\ \cite{Arguelles_review:in_prep}. 

\smallskip

  {\bf Energy spectrum:}  The spectrum of neutrinos depends on their production processes, but BSM effects could introduce identifiable features, {\it e.g.}, peaks, troughs, and cut-offs.  Present neutrino telescopes reconstruct the energy $E$ of detected events to within 0.1 in $\log_{10} (E/{\rm GeV})$\ \cite{Aartsen:2013vja}. For TeV--PeV astrophysical neutrinos, the spectrum is predicted to be a featureless power law.  IceCube data are consistent with that, but also with a broken power law\ \cite{Murase:2013rfa, Chen:2013dza, Chen:2014gxa, Aartsen:2015knd, Anchordoqui:2016ewn, Vincent:2016nut}.  For EeV cosmogenic neutrinos, the spectrum has a different but predictable shape\ \cite{Beresinsky:1969qj, Berezinsky:1975zz, Stecker:1978ah, Hill:1983xs, Yoshida:1993pt, Engel:2001hd, Anchordoqui:2007fi, Takami:2007pp, Ahlers:2009rf, Ahlers:2010fw, Kotera:2010yn, Yoshida:2012gf, Ahlers:2012rz, Aloisio:2015ega, Heinze:2015hhp, Romero-Wolf:2017xqe, AlvesBatista:2018zui, Moller:2018isk, vanVliet:2019nse, Heinze:2019jou}, so BSM effects, {\it e.g.}, modifications of neutrino-nucleon cross sections\ \cite{Kusenko:2001gj,AlvarezMuniz:2001mk,Anchordoqui:2001cg,Cornet:2001gy,Kowalski:2002gb,AlvarezMuniz:2002ga,Anchordoqui:2005pn,Marfatia:2015hva,Ellis:2016dgb,Anchordoqui:2018qom}, may also be apparent.  
 
\smallskip
 
  {\bf Arrival directions:}  If the diffuse flux of cosmic neutrinos comes from an isotropic distribution of sources, then it should be isotropic itself.  However, interactions with cosmic backgrounds might induce anisotropies. For instance, they could create a neutrino horizon, whereby high-energy neutrinos could only reach us from a few nearby sources\ \cite{Ioka:2014kca, Ng:2014pca,Cherry:2014xra}. Similarly, neutrino interactions with dark matter could introduce an anisotropy towards the Galactic Center\ \cite{Arguelles:2017atb}.  Presently, the pointing resolution at neutrino telescopes is sub-degree for events initiated by $\nu_\mu$ --- tracks --- and of a few degrees for events initiated mainly by $\nu_e$ and $\nu_\tau$ --- showers\ \cite{Aartsen:2013vja}.
 
\smallskip

  {\bf Flavor composition:}  At the neutrino sources, high-energy cosmic neutrinos are believed to be produced in the decay of pions, {\it i.e.}, $\pi^+ \to \mu^+ \nu_\mu$ followed by $\mu^+ \to e^+ \nu_e \bar{\nu}_\mu$.  This results in an initial flavor composition of $\left( \nu_e : \nu_\mu : \nu_\tau \right) = \left( 1 : 2 : 0 \right)$, adding $\nu$ and $\bar{\nu}$.  Upon reaching Earth, oscillations have transformed this into nearly $\left( 1 : 1: 1 \right)_\oplus$\ \cite{Pakvasa:2008nx}. The detection of $\nu_\tau$ is minimally required for testing this standard oscillation scenario \cite{Palladino:2018qgi,Parke:2015goa}.  
  While there are variations on this canonical expectation\ \cite{Barenboim:2003jm, Kashti:2005qa, Lipari:2007su}, the expected flavor ratios fall within a well-defined region\ \cite{Bustamante:2015waa}.  However, numerous BSM models active during propagation may modify this\ \cite{Mehta:2011qb, Rasmussen:2017ert}, including neutrino decay and Lorentz invariance violation, as shown in Fig.~\ref{fig:flavor_ratios}.   A precise measurement of the flavor composition could distinguish between these two classes of models\ \cite{Bustamante:2015waa}.  Presently, measuring flavor at neutrino telescopes is challenging, since the showers made by $\nu_e$ and $\nu_\tau$ look similar\ \cite{Aartsen:2015ivb, Aartsen:2015knd}, which makes the contours of allowed flavor composition in Fig.\ \ref{fig:flavor_ratios} wide.
 
\smallskip

  {\bf Timing:}  A violation of Lorentz invariance would modify the energy-momentum relation of neutrinos and photons\ \cite{AmelinoCamelia:2003ex, Christian:2004xb, Diaz:2014yva}, causing them to have different speeds at different energies.  This would manifest in neutrinos\ \cite{Diaz:2016xpw, Murase:2019xqi}, photons\ \cite{Longo:1987ub, Wang:2016lne, Wei:2016ygk, Boran:2018ypz}, and gravitational waves\ \cite{Baret:2011tk} emitted at the same time from transient sources arriving at Earth at different times.  Presently, electronics in neutrino telescopes can timestamp events to within a few nanoseconds\ \cite{Aartsen:2016nxy}.

\smallskip
Today, the strength of the tests performed using these observables is limited at PeV energies, where data is scant, but event statistics are growing and there are ongoing efforts to improve the reconstruction of neutrino properties.  Once neutrinos of higher energies are detected, the same observables can be used to test fundamental physics in a new energy regime.

%%%%%%%%%%%%%%%%%%%%%%%%%%%%%
\begin{figure}[t!]
 \begin{center}
  \includegraphics[width=\textwidth]{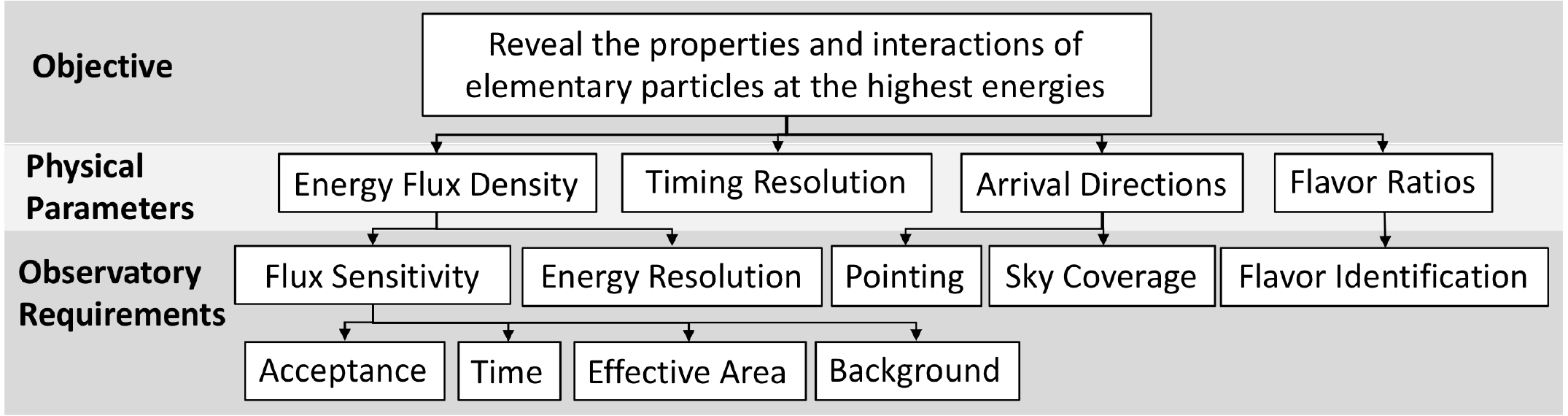}
 \end{center}
  \vspace*{-0.5 cm}
 \caption{\label{fig:physics_flow_chart}{Observatory requirements to test fundamental physics with cosmic neutrinos.}}
\end{figure}
%%%%%%%%%%%%%%%%%%%%%%%%%%%%%

\subsection*{Observatory Requirements to Achieve the Science Goals}

Answering fundamental physics questions requires improving the precision with which neutrino observables are measured, which is currently limited by the low numbers of events. The statistics in the TeV--PeV energy range will grow using existing neutrino detectors and their planned upgrades. This will be supplemented by improved techniques to reconstruct neutrino energy, direction, and flavor. At the EeV scale, our ability to address fundamental physics questions is contingent on the discovery of neutrinos at these energies. In addition to emphasizing the importance of improved statistics, we highlight two measurements that can be improved in the coming decade: the neutrino cross section and flavor composition.

Presently, the measurement of the TeV--PeV neutrino cross section in multiple energy bins is sorely statistics-limited\ \cite{Bustamante:2017xuy}. 
In this energy range, where the measured cross section is compatible with SM expectations, large BSM deviations are unlikely.  But smaller deviations are still possible, especially close to PeV energies.  To extract the cross section, Ref.\ \cite{Bustamante:2017xuy} used about 60 shower events collected by IceCube in six years across all energies. A detector that is five times larger\ \cite{Aartsen:2014njl} would collect 300 showers in the same time, reducing the statistical error in the extracted cross sections by a factor of $\sqrt{6}/6 \approx 0.4$~\cite{Anchordoqui:2019ufu}.  At that point, the statistical and systematic errors would become comparable, with a size of about 0.2 in the logarithm of the cross section (in units of cm$^2$).

At the EeV scale, measuring the cross section to within an order of magnitude could distinguish between SM predictions and BSM modifications; see Fig.\ \ref{fig:cross_section}. This target is achievable with tens of events in the PeV--EeV energy range. 
Detection will be challenging, since the flux is expected to decrease fast with energy and the cross section is expected to grow with energy, making the Earth opaque to neutrinos.
Facing significant uncertainties in the predicted flux of cosmogenic neutrinos\ \cite{Kotera:2010yn, Ahlers:2012rz, Romero-Wolf:2017xqe, AlvesBatista:2018zui, Heinze:2019jou}, we advocate for the construction of larger neutrino observatories to boost the chances of discovering and collecting a sufficiently large number of cosmogenic neutrinos.

Flavor composition must be measured with a precision better than 40\% to match the theoretical SM uncertainty band and identify BSM deviations, as shown in Fig.\ \ref{fig:flavor_ratios}. 
Reaching this target at TeV--PeV energies requires supplementing the larger event statistics with the detection of flavor-specific signals\ \cite{Glashow:1960zz, Learned:1994wg, Li:2016kra}. 
With 20\% precision, we could distinguish between models similar to neutrino decay or to Lorentz invariance violation. Improved statistics will also permit searches for a potential energy dependence of mixing, which could point to the presence of BSM effects\ \cite{Mehta:2011qb, Bustamante:2015waa}.  

In the EeV range, we advocate exploring new methods to measure flavor in existing and upcoming experiments ({\it e.g.}, Ref.\ \cite{Wang:2013bvz}).  Some planned EeV detectors will be sensitive primarily\ \cite{Fargion:1999se} to $\nu_\tau$\cite{Olinto:2017xbi, Sasaki:2017zwd, BEACON_ARENA2018:Talk, Liu:2018hux,  Alvarez-Muniz:2018bhp, Otte:2018uxj}, while others will be sensitive to all flavors~\cite{Allison:2011wk, Barwick:2014rca, Aartsen:2014njl, Gorham:2008dv, Adrian-Martinez:2016fdl, James:2017pvr}, but might not be able to distinguish between them easily.  Thus, we should consider combining data from the two types of experiments in order to infer at least the $\nu_\tau$ fraction.

Further, with the available sub-degree pointing resolution,
we can begin to probe anisotropies in the neutrino sky that may result, {\it e.g.}, from Lorentz-invariance violation\ \cite{Abbasi:2010kx} or BSM matter interactions\ \cite{Arguelles:2017atb}. Additionally, we can cull a set of neutrino events that are truly extragalactic, by using only those that point away from the Galactic Center, which allows us to make robust searches for BSM effects that are enhanced over cosmological distances ({\it e.g.}, Ref.\ \cite{Bustamante:2016ciw}). 

We advocate for a strategy for the coming decade that improves precision on flavor identification and improves statistics across a broad energy scale, from 10 TeV up to the EeV scale. 
While this strategy targets mainly cross section and flavor measurements, it will impact other neutrino observables and relentlessly test the predictions of the SM and of many BSM scenarios.

% \bibliographystyle{utphysmod.bst}
% \bibliography{references}

\providecommand{\href}[2]{#2}\begingroup\raggedright\endgroup

\end{document}

%% file: endorsers.tex
 \newcounter{AffiliationCount} 
 \def\affiliations{} 

 \newcommand{\newAff}[2]
 {
     \stepcounter{AffiliationCount}
     \xdef#1{$^{\theAffiliationCount}$}
     \appto\affiliations {{#1\textit{\mbox{#2}\quad}} }
 }

\def\affiliations{\footnotesize}

\newAff{\atUCI}{University of California, Irvine} 
\newAff{\atIOPB}{Institute of Physics, Bhubaneswar} 
\newAff{\atULB}{Universit\'e Libre de Bruxelles} 
\newAff{\atClemson}{Clemson University} 
\newAff{\atINFN}{Istituto Nazional di Fisica Nucleare (INFN)}
\newAff{\atGSSI}{Gran Sasso Science Institute (GSSI)}
\newAff{\atCompostela}{Universidade de Santiago de Compostela} 
\newAff{\atSP}{Universidade de S\~ao Paulo} 
\newAff{\atChungbuk}{Chungbuk National University}
\newAff{\atMarquette}{Marquette University}
\newAff{\atAmsterdam}{Universiteit van Amsterdam} 
\newAff{\atErlangen}{Friedrich-Alexander-Universit\"at Erlangen-N\"urnberg} 
\newAff{\atSorbonne}{Sorbonne Universit\'e} 
\newAff{\atBerne}{Universit\'e de Berne}  
\newAff{\atTokyo}{University of Tokyo} 
\newAff{\atNBI}{Niels Bohr Institute, University of Copenhagen}
\newAff{\atRWTHAachen}{Rheinisch-Westf\"alische Technische Hochschule Aachen}
\newAff{\atPSU}{Pennsylvania State University}
\newAff{\atSDSM}{South Dakota School of Mines and Technology}
\newAff{\atValencia}{Institut de F\'isica Corpuscular, Universitat de Val\`encia} 
\newAff{\atUW}{University of Wisconsin, Madison}
\newAff{\atFlorida}{University of Florida}
\newAff{\atOSU}{The Ohio State University}
\newAff{\atMelbourne}{University of Melbourne}
\newAff{\atAdelaide}{University of Adelaide}
\newAff{\atRochester}{University of Rochester}
\newAff{\atUtah}{University of Utah}
\newAff{\atPadova}{Universit\`a degli Studi di Padova}
\newAff{\atDESYZeuthen}{Deutsches Elektronen-Synchrotron (DESY) Zeuthen}
\newAff{\atTurin}{Universit\`a degli Studi di Torino}
\newAff{\atNASAMarshall}{NASA Marshall Space Flight Center}
\newAff{\atINFNTurin}{Istituto Nazional di Fisica Nucleare (INFN), Sezione di Torino}
\newAff{\atIPNO}{Institut de Physique Nucl\'eaire d'Orsay (IPNO), Universit\'e Paris-Sud, Universit\'e Paris-Saclay}
\newAff{\atUMD}{University of Maryland, College Park} 
\newAff{\atINNPPP}{Institut National de Physique Nucl\'eaire et de Physique des Particules (IN2P3)}
\newAff{\atZagreb}{University of Zagreb}
\newAff{\atUppsala}{Uppsala Universitet}
\newAff{\atCFA}{Center for Astrophysics, Harvard \& Smithsonian}
\newAff{\atINFNBari}{Istituto Nazional di Fisica Nucleare (INFN), Sezione di Bari}
\newAff{\atGoddard}{NASA Goddard Space Flight Center}
\newAff{\atCatania}{Universit\`a degli Studi di Catania}
\newAff{\atFSU}{Florida State University}
\newAff{\atMPI}{Max-Planck-Institut f\"ur Kernphysik, Heidelberg} 
\newAff{\atWitwatersrand}{University of the Witwatersrand}
\newAff{\atNTU}{National Taiwan University}
\newAff{\atAnnaba}{Badji Mokhtar University of Annaba}
\newAff{\atCUNY}{City University of New York}
\newAff{\atVUB}{Vrije Universiteit Brussels}
\newAff{\atMarseille}{Centre de Physique des Particules de Marseille (CPPM)}
\newAff{\atUCL}{University College London}
\newAff{\atHongKong}{The University of Hong Kong}
\newAff{\atTata}{Tata Institute of Fundamental Research, Mumbai (TIFR)}
\newAff{\atNW}{Northwestern University}
\newAff{\atRio}{Universidade Federal do Rio de Janeiro}
\newAff{\atNijmegen}{Radboud Universiteit Nijmegen}
\newAff{\atNikhef}{Nikhef}
\newAff{\atIAP}{Institut d'Astrophysique de Paris}
\newAff{\atBNL}{Brookhaven National Laboratory} 
\newAff{\atMSU}{Michigan State University}
\newAff{\atChicago}{University of Chicago}
\newAff{\atMunster}{Westf\"alische Wilhelms-Universit\"at M\"unster}
\newAff{\atLund}{Lunds Universitet}
\newAff{\atLMU}{Ludwig-Maximilians-Universit\"at M\"unchen}
\newAff{\atRIKEN}{RIKEN}
\onlyfundamental{\newAff{\atKings}{King's College London}}
\newAff{\atKIT}{Karlsruher Institut f\"ur Technologie}
\newAff{\atColoradoMines}{Colorado School of Mines} 
\newAff{\atPontificiaRio}{Pontificia Universidade Catolic\'a do Rio de Janeiro}
\newAff{\atStanford}{Stanford University}
\newAff{\atOKState}{Oklahoma State University} 
\newAff{\atStockholm}{Stockholm Universitet}
\newAff{\atKyoto}{Kyoto University}
\newAff{\atUAM}{Instituto de F\'isica Te\'orica UAM-CSIC}
\newAff{\atPeru}{Pontificia Universidad Cat\'olica del Per\'u}
\newAff{\atDelaware}{Bartol Research Institute, University of Delaware}
\newAff{\atLaPlata}{Universidad Nacional de La Plata}
\newAff{\atTuebingen}{Eberhard Karls Universit\"at T\"ubingen} 
\newAff{\atFirenze}{Universit\`a degli Studi di Firenze} 
\newAff{\atUCLA}{University of California, Los Angeles}
\newAff{\atWhittier}{Whittier College}
\newAff{\atUNF}{University of North Florida}
\newAff{\atFermilab}{Fermi National Accelerator Laboratory}
\newAff{\atVT}{Virginia Polytechnic Institute and State University} 
\newAff{\atIPHC}{Institut Pluridisciplinaire Hubert Curien (IPHC)}
\newAff{\atSaitama}{Saitama University}
\newAff{\atICTPSAIFR}{International Center for Theoretical Physics -- South American Institute for Fundamental Research} 
\newAff{\atICRAR}{International Centre for Radio Astronomy Research, Curtin University}
\newAff{\atBerkeley}{University of California, Berkeley} 
\onlyastrophysical{\newAff{\atWurzburg}{Julius-Maximilians-Universit\"at W\"urzburg}}
\newAff{\atKonan}{Konan University}
\newAff{\atJHU}{Johns Hopkins University}
\newAff{\atQM}{Queen Mary University of London}
\newAff{\atColumbia}{Columbia University} 
\newAff{\atChiba}{Chiba University}
\newAff{\atLBL}{Lawrence Berkeley National Laboratory}
\newAff{\atHumboldt}{Humboldt-Universit\"at zu Berlin} 
\newAff{\atGutenberg}{Johannes Gutenberg-Universit\"at Mainz}
\newAff{\atCERN}{CERN} 
\onlyfundamental{\newAff{\atIndiana}{Indiana University}}
\newAff{\atAPC}{Laboratoire AstroParticule et Cosmologie} 
\newAff{\atNebraska}{University of Nebraska-Lincoln}
\newAff{\atUMBC}{University of Maryland, Baltimore County}
\newAff{\atDrexel}{Drexel University}
\newAff{\atINFNPisa}{Istituto Nazional di Fisica Nucleare (INFN), Sezione di Pisa}
\newAff{\atUAH}{University of Alabama in Huntsville}
\newAff{\atKU}{University of Kansas}
\newAff{\atHawaii}{University of Hawaii, Manoa}
\newAff{\atMIT}{Massachusetts Institute of Technology} 
\newAff{\atSLAC}{SLAC National Accelerator Lab}
\newAff{\atTrieste}{Universit\`a degli Studi di Trieste}
\newAff{\atASU}{Arizona State University}
\newAff{\atSapienza}{Sapienza – Universit\`a di Roma}
\newAff{\atUWRF}{University of Wisconsin-River Falls}
\newAff{\atINFNPadova}{Istituto Nazional di Fisica Nucleare (INFN), Sezione di Padova}
\onlyfundamental{\newAff{\atWurzburg}{Julius-Maximilians-Universit\"at W\"urzburg}}
\newAff{\atPuebla}{Benem\'erita Universidad Aut\'onoma de Puebla} 
\newAff{\atGranada}{Universidad de Granada}
\newAff{\atQueens}{Queen's University}
\onlyastrophysical{\newAff{\atKings}{King's College London}}
\newAff{\atMercer}{Mercer University}
\onlyfundamental{\newAff{\atCalPoly}{California Polytechnic State University}}
\newAff{\atMilan}{Universit\`a degli Studi di Milano}
\newAff{\atCinestav}{Centro de Investigaci\'on y de Estudios Avanzados del Instituto Polit\'ecnico Nacional (Cinvestav)}
\newAff{\atMPIMunich}{Max-Planck-Institut fur Physik, M\"unchen}
\newAff{\atGeneva}{Universit\'e de Gen\`eve}
\newAff{\atAlberta}{University of Alberta}
\newAff{\atWeizmann}{Weizmann Institute of Science} 
\newAff{\atAuvergne}{Universit\'e Clermont Auvergne}
\newAff{\atMTU}{Michigan Technological University} 
\newAff{\atMontpellier}{Universit\'e de Montpellier}
\newAff{\atESO}{European Southern Observatory} 
\newAff{\atGT}{Georgia Institute of Technology} 
\newAff{\atCrete}{University of Crete} 
\newAff{\atTorVergata}{Universit\`a degli Studi di Roma Tor Vergata}
\newAff{\atNCBJ}{Naradowe Centrum Bada\'n J\k{a}drowych}
\newAff{\atISSRomania}{Institutul de \cb{S}tiin\cb{t}e Spa\cb{t}iale } 
\newAff{\atWashU}{Washington University in St. Louis}
\newAff{\atJoburg}{University of Johannesburg} 
\onlyastrophysical{\newAff{\atLUTH}{Laboratoire Univers et Th\'eories}  }
\newAff{\atIowa}{University of Iowa} 
\newAff{\atTUM}{Technische Universit\"at M\"unchen} 
\newAff{\atUCSD}{University of California, San Diego}  
\newAff{\atVUAms}{Vrije Universiteit Amsterdam} 
\newAff{\atSKKU}{Sungkyunkwan University (SKKU)}
\newAff{\atMichigan}{University of Michigan, Ann Arbor} 
\newAff{\atLeiden}{Universiteit Leiden}
\newAff{\atBarca}{Universitat de Barcelona} 
\newAff{\atBama}{University of Alabama} 
\newAff{\atWuppertal}{Bergische Universit\"at Wuppertal}
\newAff{\atGroningen}{Rijksuniversiteit Groningen} 
\onlyastrophysical{\newAff{\atSaclay}{Universit\'e Paris-Saclay}}
\newAff{\atNapoli}{Universit\`a degli Studi di Napoli Federico II}
\newAff{\atAnnecy}{Universit\'e Grenoble Alpes, Laboratoire d'Annecy-le-Vieux de Physique Th\'eorique (LAPTh)} 
\newAff{\atHamburg}{Universit\"at Hamburg} 
\newAff{\atPrinceton}{Princeton University}
\newAff{\atBologna}{Universit\`a degli Studi di Bologna}
\newAff{\atGenova}{Universit\`a degli Studi di Genova}
\newAff{\atINFNGenova}{Istituto Nazional di Fisica Nucleare (INFN), Sezione di Genova}
\newAff{\atISSCSIC}{Institute of Space Sciences (IEEC-CSIC)}
\newAff{\atTufts}{Tufts University}
\newAff{\atLaAquila}{Universit\`a degli Studi dell'Aquila}
\newAff{\atLeeds}{University of Leeds}
\newAff{\atvandy}{Vanderbilt University}
% \onlyastrophysical{\newAff{\atUCLA}{University of California, Los Angeles}}
\newAff{\atTDLI}{Tsung-Dao Lee Institute}
\newAff{\atSunYetsen}{Sun Yet-sen University}
\newAff{\atNovaGorica}{Univerza v Novi Gorici}
\newAff{\atUNLV}{University of Nevada, Las Vegas}
\newAff{\atLosAlamos}{Los Alamos National Laboratory}
\newAff{\atInnsbruck}{Leopold-Franzens-Universit\"at Innsbruck}

\mbox{Kevork N. Abazajian}\atUCI,
\mbox{Sanjib Kumar Agarwalla}\atIOPB,
\mbox{Juan Antonio Aguilar S\'anchez}\atULB,
\mbox{Marco Ajello}\atClemson,
\mbox{Roberto Aloisio}\atINFN~\atGSSI, 
\mbox{Jaime \'Alvarez-Mu\~niz}\atCompostela,
\mbox{Rafael Alves Batista}\atSP,
\mbox{Hongjun An}\atChungbuk,
\mbox{Karen Andeen}\atMarquette, 
\mbox{Shin'ichiro Ando}\atAmsterdam,
\mbox{Gisela Anton}\atErlangen,
\mbox{Ignatios Antoniadis}\atSorbonne~\atBerne,
\mbox{Katsuaki Asano}\atTokyo,
\mbox{Katie Auchettl}\atNBI,
\mbox{Jan Auffenberg}\atRWTHAachen,
\mbox{Hugo Ayala}\atPSU,
\mbox{Xinhua Bai}\atSDSM,
\mbox{Gabriela Barenboim}\atValencia,
\mbox{Vernon Barger}\atUW,
\mbox{Imre Bartos}\atFlorida,
\mbox{Steve W. Barwick}\atUCI,
\mbox{John Beacom}\atOSU,
\mbox{James J. Beatty}\atOSU,
\mbox{Nicole F. Bell}\atMelbourne,
\mbox{Jos\'e Bellido}\atAdelaide,
\mbox{Segev BenZvi}\atRochester,
\mbox{Douglas R. Bergman}\atUtah,
\mbox{Jos\'e Bernab\'eu}\atValencia,
\mbox{Elisa Bernardini}\atPadova~\atDESYZeuthen,
\mbox{Mario Bertaina}\atTurin,
\mbox{Gianfranco Bertone}\atAmsterdam,
\mbox{Peter F. Bertone}\atNASAMarshall,
\mbox{Francesca Bisconti}\atINFNTurin,
\mbox{Jonathan Biteau}\atIPNO,
\mbox{Erik Blaufuss}\atUMD,
\mbox{Summer Blot}\atDESYZeuthen,
\mbox{Julien Bolmont}\atINNPPP,
\mbox{Zeljka Bosnjak}\atZagreb,
\mbox{Olga Botner}\atUppsala,
\mbox{Federica Bradascio}\atDESYZeuthen,
\mbox{Esra Bulbul}\atCFA,
\mbox{Alexander Burgman}\atUppsala,
\mbox{Francesco Cafagna}\atINFNBari,
\mbox{Regina Caputo}\atGoddard,
\mbox{M. Carmen Carmona-Benitez}\atPSU,
\mbox{Rossella Caruso}\atCatania,
\mbox{Marco Casolino}\atINFN,
\mbox{Karem Pe\~nal\'o Castillo}\atFSU,
\mbox{Silvia Celli}\atMPI,
\onlyastrophysical{\mbox{S. Bradley Cenko}\atGoddard,}
\mbox{Andrew Chen}\atWitwatersrand,
\mbox{Yaocheng Chen}\atNTU,
\mbox{Talai Mohamed Cherif}\atAnnaba,
\mbox{Nafis Rezwan Khan Chowdhury}\atValencia,
\mbox{Eugene M. Chudnovsky}\atCUNY,
\mbox{Brian A. Clark}\atOSU,
\mbox{Pablo Correa}\atVUB,
\mbox{Doug F. Cowen}\atPSU,
\mbox{Paschal Coyle}\atMarseille, 
\mbox{Linda Cremonesi}\atUCL,
\mbox{Jane Lixin Dai}\atHongKong, 
\mbox{Basudeb Dasgupta}\atTata,
\mbox{Andr\'e de Gouv\^ea}\atNW,
\mbox{Sijbrand de Jong}\atNijmegen~\atNikhef,
\mbox{Simon De Kockere}\atVUB,
\mbox{Jo\~ao R. T. de Mello Neto}\atRio,
\mbox{Luiz de Viveiros}\atPSU,
\mbox{Krijn D. de Vries}\atVUB,
\onlyastrophysical{\mbox{Gwenha\"el de Wasseige}\atAPC,}
\mbox{Valentin Decoene}\atIAP,
\mbox{Peter B. Denton}\atBNL,
\mbox{Tyce DeYoung}\atMSU,
\mbox{Rebecca Diesing}\atChicago,
\mbox{Markus Dittmer}\atMunster,
\mbox{Caterina Doglioni}\atLund,
\mbox{Klaus Dolag}\atLMU,
\mbox{Michele Doro}\atPadova,
\mbox{Michael A. DuVernois}\atUW,
\mbox{Toshikazu Ebisuzaki}\atRIKEN,
\onlyfundamental{\mbox{John Ellis}\atKings,}
\mbox{Rikard Enberg}\atUppsala, 
\mbox{Ralph Engel}\atKIT,
\mbox{Johannes Eser}\atColoradoMines,
\mbox{Arman Esmaili}\atPontificiaRio,
\mbox{Ke Fang}\atStanford,
\mbox{Jonathan L. Feng}\atUCI,
\mbox{Gustavo Figueiredo}\atOKState,
\mbox{George Filippatos}\atColoradoMines,
\mbox{Chad Finley}\atStockholm,
\mbox{Derek Fox}\atPSU,
\mbox{Anna Franckowiak}\atDESYZeuthen,
\mbox{Elizabeth Friedman}\atUMD,
\mbox{Toshihiro Fujii}\atKyoto,
\mbox{Daniele Gaggero}\atUAM,
\mbox{Alberto M. Gago}\atPeru,
\mbox{Thomas Gaisser}\atDelaware,
\mbox{Shan Gao}\atDESYZeuthen,
\mbox{Carlos Garc\'ia Canal}\atLaPlata,
\mbox{Daniel Garc\'ia-Fern\'andez}\atDESYZeuthen,
\mbox{Simone Garrappa}\atDESYZeuthen,
\mbox{Maria Vittoria Garzelli}\atTuebingen~\atFirenze,
\mbox{Graciela B. Gelmini}\atUCLA,
\mbox{Christian Glaser}\atUCI,
\mbox{Allan Hallgren}\atUppsala,
\mbox{Jordan C. Hanson}\atWhittier,
\mbox{Andreas Haungs}\atKIT,
\mbox{John W. Hewitt}\atUNF,
\mbox{Jannik Hofest\"adt}\atErlangen,
\mbox{Kara Hoffman}\atUMD,
\mbox{Benjamin Hokanson-Fasig}\atUW,
\mbox{Dan Hooper}\atFermilab~\atChicago,
\mbox{Shunsaku Horiuchi}\atVT,
\mbox{Feifei Huang}\atIPHC,
\mbox{Patrick Huber}\atVT, 
\mbox{Tim Huege}\atKIT,
\mbox{Kaeli Hughes}\atChicago,
\mbox{Naoya Inoue}\atSaitama,
\mbox{Susumu Inoue}\atRIKEN,
\mbox{Fabio Iocco}\atICTPSAIFR,
\mbox{Kunihito Ioka}\atKyoto,
\mbox{Clancy W. James}\atICRAR,
\mbox{Eleanor Judd}\atBerkeley,
\mbox{Daniel Kabat}\atCUNY,
\onlyastrophysical{\mbox{Matthias Kadler}\atWurzburg,}
\mbox{Fumiyoshi Kajino}\atKonan,
\mbox{Takaaki Kajita}\atTokyo,
\mbox{Marc Kamionkowski}\atJHU,
\mbox{Alexander Kappes}\atMunster,
\mbox{Dimitra Karabali}\atCUNY,
\mbox{Timo Karg}\atDESYZeuthen,
\mbox{Teppei Katori}\atQM,
\mbox{Uli F. Katz}\atErlangen,
\onlyfundamental{\mbox{Norita Kawanaka}\atKyoto,}
\mbox{Azadeh Keivani}\atColumbia,
\mbox{John L. Kelley}\atUW,
\mbox{Myoungchul Kim}\atChiba,
\mbox{Shigeo S. Kimura}\atPSU,
\mbox{Spencer Klein}\atLBL,
\mbox{Stefan Klepser}\atDESYZeuthen, 
\mbox{David Koke}\atMunster,
\mbox{Hermann Kolanoski}\atHumboldt, 
\mbox{Lutz K\"opke}\atGutenberg,
\mbox{Joachim Kopp}\atGutenberg~\atCERN,
\mbox{Claudio Kopper}\atMSU,
\mbox{Jason Koskinen}\atNBI,
\onlyfundamental{\mbox{V. Alan Kosteleck\'{y}}\atIndiana,}
\mbox{Dmitriy Kostunin}\atDESYZeuthen,
\mbox{Antoine Kouchner}\atAPC,
\mbox{Ilya Kravchenko}\atNebraska,
\mbox{John Krizmanic}\atUMBC,
\mbox{Naoko Kurahashi Neilson}\atDrexel,
\mbox{Michael Kuss}\atINFNPisa,
\mbox{Evgeny Kuznetsov}\atUAH,
\mbox{Ranjan Laha}\atCERN,
\mbox{Uzair Abdul Latif}\atKU,
\mbox{John G. Learned}\atHawaii,
\mbox{Jean-Philippe Lenain}\atSorbonne,
\mbox{Rebecca K. Leane}\atMIT,
\mbox{Shirley Weishi Li}\atSLAC,
\mbox{Lu Lu}\atChiba,
\mbox{Francesco Longo}\atTrieste,
\mbox{Andrew Ludwig}\atChicago,
\mbox{Cecilia Lunardini}\atASU,
\mbox{Paolo Lipari}\atSapienza,
\mbox{James Madsen}\atUWRF,
\mbox{Keiichi Mase}\atChiba,
\mbox{Manuela Mallamaci}\atINFNPadova,
\mbox{Karl Mannheim}\atWurzburg,
\mbox{Danny Marfatia}\atHawaii,
\mbox{Raffaella Margutti}\atNW,
\mbox{Cristian Jes\'us Lozano Mariscal}\atMunster,
\mbox{Szabolcs Marka}\atColumbia,
\mbox{Olivier Martineau-Huynh}\atINNPPP,
\mbox{Oscar Mart\'inez-Bravo}\atPuebla,
\mbox{Manuel Masip}\atGranada,
\mbox{Nikolaos E. Mavromatos}\atKings,
\mbox{Arthur B. McDonald}\atQueens,
\mbox{Frank McNally}\atMercer,
\mbox{Olga Mena}\atValencia,
\mbox{Kevin-Druis Merenda}\atColoradoMines,
\mbox{Philipp Mertsch}\atRWTHAachen,
\mbox{Peter M\'esz\'aros}\atPSU,
\onlyfundamental{\mbox{Matthew Mewes}\atCalPoly,}
\mbox{Hisakazu Minakata}\atTokyo,
\mbox{Nestor Mirabal}\atGoddard,
\mbox{Lino Miramonti}\atMilan,
\mbox{Omar G. Miranda}\atCinestav,
\mbox{Razmik Mirzoyan}\atMPIMunich,
\mbox{John W. Mitchell}\atGoddard,
\mbox{Irina Mocioiu}\atPSU,
\mbox{Teresa Montaruli}\atGeneva,
\mbox{Maria Elena Monzani}\atSLAC,
\mbox{Roger Moore}\atAlberta,
\mbox{Shigehiro Nagataki}\atRIKEN,
\mbox{Masayuki Nakahata}\atTokyo,
\mbox{Jiwoo Nam}\atNTU,
\mbox{Kenny C. Y. Ng}\atWeizmann,
\mbox{Ryan Nichol}\atUCL,
\mbox{Valentin Niess}\atAuvergne,
\mbox{David F. Nitz}\atMTU,
\mbox{Samaya Nissanke}\atAmsterdam,
\mbox{Eric Nuss}\atMontpellier,
\mbox{Eric Oberla}\atChicago,
\mbox{Stefan Ohm}\atDESYZeuthen,
\mbox{Kouji Ohta}\atKyoto,
\mbox{Foteini Oikonomou}\atESO,
\mbox{Roopesh Ojha}\atUMBC~\atGoddard,
\mbox{Nepomuk Otte}\atGT,
\mbox{Timothy A. D. Paglione}\atCUNY,
\mbox{Sandip Pakvasa}\atHawaii,
\mbox{Andrea Palladino}\atDESYZeuthen,
\mbox{Sergio Palomares-Ruiz}\atValencia,
\mbox{Vasiliki Pavlidou}\atCrete,
\mbox{Carlos P\'erez de los Heros}\atUppsala,
\mbox{Christopher Persichilli}\atUCI,
\mbox{Piergiorgio Picozza}\atINFN~\atTorVergata,
\mbox{Zbigniew Plebaniak}\atNCBJ,
\mbox{Vlad Popa}\atISSRomania,
\mbox{Steven Prohira}\atOSU,
\mbox{Bindu Rani}\atGoddard,
\mbox{Brian Flint Rauch}\atWashU,
\mbox{Soebur Razzaque}\atJoburg,
\onlyastrophysical{\mbox{Nicolas Renault-Tinacci}\atLUTH,}
\mbox{Mary Hall Reno}\atIowa,
\mbox{Elisa Resconi}\atTUM,
\mbox{Marco Ricci}\atINFN,
\mbox{Jarred M. Roberts}\atUCSD,
\mbox{Nicholas L. Rodd}\atBerkeley~\atLBL,
\onlyfundamental{\mbox{Werner Rodejohann}\atMPI,}
\mbox{Juan Rojo}\atVUAms,
\mbox{Carsten Rott}\atSKKU,
\mbox{Iftach Sadeh}\atDESYZeuthen,
\mbox{Benjamin R. Safdi}\atMichigan,
\mbox{Naoto Sakaki}\atRIKEN,
\onlyfundamental{\mbox{David Saltzberg}\atUCLA,}
\mbox{Jordi Salvad\'o}\atBarca,
\mbox{Dorothea Samtleben}\atLeiden,
\mbox{Marcos Santander}\atBama,
\mbox{Fred Sarazin}\atColoradoMines,
\mbox{Konstancja Satalecka}\atDESYZeuthen,
\mbox{Michael Schimp}\atWuppertal,
\mbox{Olaf Scholten}\atGroningen,
\mbox{Harm Schoorlemmer}\atMPI,
\onlyastrophysical{\mbox{Frank G. Schr\"oder}\atDelaware,}
\onlyastrophysical{\mbox{Fabian Sch\"ussler}\atSaclay,}
\mbox{Sergio J. Sciutto}\atLaPlata,
\mbox{Valentina Scotti}\atNapoli,
\mbox{David Seckel}\atDelaware,
\mbox{Pasquale D. Serpico}\atAnnecy,
\mbox{Shashank Shalgar}\atNBI,
\mbox{Jerry Shiao}\atNTU,
\onlyastrophysical{\mbox{Ankur Sharma}\atINFNPisa,}
\mbox{Kenji Shinozaki}\atTurin,
\mbox{Ian M. Shoemaker}\atVT,
\mbox{G\"unter Sigl}\atHamburg,
\mbox{Lorenzo Sironi}\atColumbia,
\mbox{Tracy R. Slatyer}\atMIT,
\mbox{Radomir Smida}\atChicago,
\mbox{Alexei Yu Smirnov}\atMPI,
\mbox{Jorge F. Soriano}\atCUNY,
\mbox{Daniel Southall}\atChicago,
\mbox{Glenn Spiczak}\atUWRF,
\mbox{Anatoly Spitkovsky}\atPrinceton,
\mbox{Maurizio Spurio}\atBologna,
\mbox{Juliana Stachurska}\atDESYZeuthen, 
\mbox{Krzysztof Z. Stanek}\atOSU,
\mbox{Floyd Stecker}\atGoddard,
\mbox{Christian Stegmann}\atDESYZeuthen,
\mbox{Robert Stein}\atDESYZeuthen,
\mbox{Anna M. Suliga}\atNBI,
\mbox{Greg Sullivan}\atUMD,
\mbox{Jacek Szabelski}\atNCBJ,
\onlyfundamental{\mbox{Ignacio Taboada}\atGT,} 
\mbox{Yoshiyuki Takizawa}\atRIKEN,
\mbox{Mauro Taiuti}\atGenova~\atINFNGenova,
\mbox{Irene Tamborra}\atNBI,
\mbox{Xerxes Tata}\atHawaii,
\mbox{Todd A. Thompson}\atOSU,
\mbox{Charles Timmermans}\atNijmegen~\atNikhef,
\mbox{Kirsten Tollefson}\atMSU,
\mbox{Diego F. Torres}\atISSCSIC,
\mbox{Jorge Torres}\atOSU,
\mbox{Simona Toscano}\atULB,
\mbox{Delia Tosi}\atUW,
\mbox{Mat\'ias Tueros}\atLaPlata,
\mbox{Sara Turriziani}\atRIKEN,
\mbox{Elisabeth Unger}\atUppsala,
\mbox{Michael Unger}\atKIT,
\mbox{Martin Unland Elorrieta}\atMunster,
\mbox{Jos\'e Wagner Furtado Valle}\atValencia,
\mbox{Lawrence Wiencke}\atColoradoMines,
\mbox{Nick van Eijndhoven}\atVUB,
\mbox{Jakob van Santen}\atDESYZeuthen,
\mbox{Arjen van Vliet}\atDESYZeuthen,
\mbox{Justin Vandenbroucke}\atUW, 
\mbox{Gary S. Varner}\atHawaii,
\mbox{Tonia Venters}\atGoddard,
\mbox{Matthias Vereecken}\atVUB,
\mbox{Alex Vilenkin}\atTufts,
\mbox{Francesco L. Villante}\atLaAquila, 
\mbox{Aaron Vincent}\atQueens,
\mbox{Martin Vollmann}\atTUM,
\mbox{Philip von Doetinchem}\atHawaii,
\mbox{Alan A. Watson}\atLeeds,
\mbox{Eli Waxman}\atWeizmann,
\mbox{Thomas Weiler}\atvandy,
\mbox{Christoph Welling}\atDESYZeuthen,
\mbox{Nathan Whitehorn}\atUCLA,
\mbox{Dawn R. Williams}\atBama,
\mbox{Walter Winter}\atDESYZeuthen,
\mbox{Hubing Xiao}\atINFNPadova,
\mbox{Donglian Xu}\atTDLI,
\mbox{Tokonatsu Yamamoto}\atKonan,
\mbox{Lili Yang}\atSunYetsen,
\mbox{Gaurang Yodh}\atUCI,
\mbox{Shigeru Yoshida}\atChiba,
\mbox{Tianlu Yuan}\atUW,
\mbox{Danilo Zavrtanik}\atNovaGorica,
\mbox{Arnulfo Zepeda}\atCinestav,
\mbox{Bing Zhang}\atUNLV,
\mbox{Hao Zhou}\atLosAlamos,
\mbox{Anne Zilles}\atIAP,
\mbox{Stephan Zimmer}\atInnsbruck,
\mbox{Juan de Dios Zornoza}\atValencia,
\mbox{Renata Zukanovich Funchal}\atSP,
\mbox{~and~Juan Z\'u\~niga}\atValencia
\linebreak

\affiliations